# *Slice emittance, projected emittance and properties of the FEL SASE radiation*

*G. Dattoli, M. Del Franco [a], A. Petralia [b], C. Ronsivalle [b] and E. Sabia [c]*

ENEA, Unità Tecnica Sviluppo di Applicazioni delle Radiazioni,
Laboratorio Modellistica Matematica, Centro Ricerche Frascati,
C.P. 65 - 00044 Frascati, Rome (Italy)

[a] ENEA Guest

[b] ENEA, Unità Tecnica Sviluppo di Applicazioni delle Radiazioni,
Laboratorio Sorgenti di Radiazioni, Centro Ricerche Frascati,
C.P. 65 - 00044 Frascati, Rome (Italy)

[c] ENEA, Unità Tecnica Tecnologie Portici, Centro Ricerche Portici,
Via Vecchio Macello - 80055 Portici, Napoli (Italy)

**ABSTRACT**

The existence of a characteristic coherence length in FEL SASE Physics determines the independent lasing of different portions, namely the slices, of the electron bunch. Each slice may be characterized by different phase space properties (not necessarily equal emittances and Twiss coefficients). This fact opens new questions on the concept of beam matching and how the various portions of the beam contribute to the performances of the output radiation, including those associated with the transverse coherence.

## I. INTRODUCTION

In this paper we will discuss the link between slice and projected emittances, the relevant transport problems, their contribution to the SASE FEL dynamics and touch on how the phase space properties of the emitted radiation are affected by the different contributions from the different slices.

The concept of slice emittance [1], and of projected emittance as well, has been introduced in SASE FEL Physics as a consequence of the fact that the output laser field is the result of the competition between different bursts of radiation emitted by different portions (the slices) of the bunch.

The longitudinal dimension of the slice is associated with the so called coherence length, given by

$$l_c = \frac{\lambda}{4\pi\sqrt{3}\rho} \qquad (1)$$

where $\lambda$ is the wavelength of the emitted radiation and $\rho$ is the gain parameter.

From the physical point of view, the coherence length corresponds to the length spanned by the radiation, in one undulator passage, during its slipping over the electron bunch. This fact ensures that the radiation, emitted by one slice, is coherent, furthermore a kind of chaotic competition occurs between the radiation, emitted at different position, in the bunch, which, being not locked in phase, grow independently.

The number of slices is simply linked to the rms bunch length $\sigma_z$ by the relation [2]

$$m \cong \frac{\sigma_z}{2\pi l_c} \qquad (2).$$

In Figure 1 we have reported a simulation with the SPARC parameters, in which we have shown the laser field longitudinal distribution, the spikes appearing in the figure are due to the competition between the various coherent portions of the radiation field, emitted by the individual independent slices

The statistical properties of the FEL SASE spikes are understood [3], albeit partially only, and will be the topic of a forthcoming investigation; here we will address the problem of understanding how the transverse coherence properties of the laser radiation are affected by the slice phase space distribution.

According to Figs. 2 we consider a bunch of electrons with a Gaussian distribution

$$l(z) = \frac{1}{\sqrt{2\pi}\,\sigma_z}\exp(-\frac{z^2}{2\sigma_z^2}) \tag{3}$$

which has been “sliced” in such a way that each segment corresponds to a coherence length, in correspondence of each slice we select a transverse region with the phase space distribution given by

$$f_n(x,x') = \frac{1}{2\pi\varepsilon_n}\exp(-\frac{\gamma_n x^2 + 2\alpha_n x x' + \beta_n x'^2}{2\varepsilon_n}) \tag{4}$$

The geometrical meaning of the various quantities appearing in eq. (4) is provided by Fig. 3. The projected phase space distributions can accordingly be defined as

$$\Phi(x,x') = \sum_{n=-N}^{N} c_n f_n(x,x') \tag{5}$$

where*

$$c_n = l(z_n)\, l_c \tag{6}$$

$z_n$ is located at the centre of each slice and it is evident that

$$\sum_n c_n = 1 \tag{7.}$$

The conclusions we can draw from the previous analysis is that

a) The projected phase-space is obtained as the weighted sum of the slice phase space contributions
b) The second moments of the projected distribution are

$$\begin{aligned} \langle x^2 \rangle &= \sum_n c_n \beta_n \varepsilon_n, \\ \langle x\, x' \rangle &= \sum_n c_n \alpha_n \varepsilon_n, \\ \langle x'^2 \rangle &= \sum_n c_n \gamma_n \varepsilon_n \end{aligned} \tag{8}$$

and, accordingly, the projected emittance and Twiss parameters are defined as

$$\begin{aligned} &\varepsilon_p = \sqrt{\langle x^2 \rangle \langle x'^2 \rangle - \langle x\, x' \rangle^2}, \\ &\gamma_p = \frac{\langle x'^2 \rangle}{\varepsilon_p}, \beta_p = \frac{\langle x^2 \rangle}{\varepsilon_p}, \alpha_p = -\frac{\langle x\, x' \rangle}{\varepsilon_p} \end{aligned} \tag{9.}$$

Examples of slice Courant-Snyder ellipses and of their projected counterpart are shown in Fig. 4, where we have chosen 5 slices, with randomly chosen (normalized) emittances up to $2\, mm \cdot mrad$.

It is evident that the projected distribution, provided by an (incoherent) superposition of Gaussian functions, is not necessarily Gaussian, the deviation from the Gaussian can be determined through the evaluation of the higher order moments, as we will discuss in the forthcoming section.

## II. RADIATION AND E-BUNCH PHASE SPACE DISTRIBUTION

Even though the problem of the slice-projected emittances has been carefully treated in the FEL literature, it seems that the effect induced by the slice components on the output SASE radiation has not received the necessary attention.

The phase space distribution of the output radiation is (at very first approximation) given by the following convolution

$$R(x,x') = \int_{-\infty}^{\infty} d\xi' \int_{-\infty}^{\infty} d\xi \, r(\xi,\xi')\,\Phi(x-\xi, x'-\xi') \tag{10}$$

Where $r(x,x')$ is the phase space radiation corresponding to the Wigner distribution of an ideal field with "emittance"[4]

$$\varepsilon_r = \frac{\lambda}{4\pi} \tag{11}$$

The derivation of the convolution integral in eq. (10) is simplified by the introduction of the formalism of quadratic forms. We define indeed the matrix

$$\hat{\Sigma} = \begin{pmatrix} \beta\varepsilon & -\alpha\varepsilon \\ -\alpha\varepsilon & \gamma\varepsilon \end{pmatrix} \tag{12}$$

and denote with $\left|\hat{\Sigma}\right|$ its determinant. The slice phase space distribution can therefore be written as

$$f_n(x,x') = \frac{1}{2\pi\sqrt{\left|\hat{\Sigma}_n\right|}} \exp\left(-\frac{1}{2}\underline{\delta}^T \hat{\Sigma}_n^{-1} \underline{\delta}\right), \qquad \underline{\delta} = \begin{pmatrix} x \\ x' \end{pmatrix} \tag{13}$$

and the use of the identity† [5]

$$\int_{-\infty}^{\infty} d^n x \exp(-\underline{x}^T \hat{A}\, \underline{x} + \underline{b}^T\, \underline{x}) = \sqrt{\frac{\pi^n}{|\hat{A}|}} \exp(-\frac{1}{4} \underline{b}^T \hat{A}^{-1} \underline{b}) \tag{14}$$

where $\underline{b}$ is an arbitrary vector independent of the integration variables, allows to write

$$R_n(x,x') = \int_{-\infty}^{\infty} d\xi' \int_{-\infty}^{\infty} d\xi\, r(\xi,\xi')\, f_n(x-\xi, x'-\xi') =$$
$$= \frac{1}{2\pi} \frac{1}{\sqrt{|\hat{T}_{n,r}|}} \exp(-\frac{1}{2} \underline{\delta}^T \hat{T}_{n,r}^{-1} \underline{\delta}), \qquad (15).$$
$$\hat{T}_{n,r}^{-1} = \hat{\Sigma}_n^{-1} - \hat{\Sigma}_n^{-1} (\hat{\Sigma}_n^{-1} + \hat{\Sigma}_r^{-1})^{-1} \hat{\Sigma}_n^{-1}$$

The convoluted emittance can be calculated from the previous relations and we find

$$\varepsilon_{n,r} = \sqrt{\varepsilon_n^2 + \varepsilon_r^2 + c_{n,r} \varepsilon_n \varepsilon_r}\,,$$
$$c_{n,r} = \gamma_r \beta_n + \gamma_n \beta_r - 2\alpha_r \beta_n \qquad (16).$$

The radiation phase-space distribution due to the contribution of all the slices is given by

$$R(x,x') = \sum_{n=-N}^{N} c_n R_n(x,x') \qquad (17).$$

However, if we assume that the total phase space distribution is provided by a Gaussian, with the Twiss coefficients given in eq. (9) we find that the convoluted radiation emittance is

$$\varepsilon_{p,r} = \sqrt{\varepsilon_p^2 + \varepsilon_r^2 + c_{p,r} \varepsilon_p \varepsilon_r} \qquad (18).$$

Let us now assume that the electron bunch consists of two slices with equal emittances and with the Courant Snyder ellipses reported in Fig. 5. Each slice is assumed to contain the same number of electrons and to satisfy the condition

$$\varepsilon_1 = \varepsilon_2 \cong \frac{\lambda}{4\pi} \tag{19},$$

By these assumptions we find

$$\varepsilon_p = \sqrt{2}\sqrt{1 + \frac{\beta_1\gamma_2 + \beta_2\gamma_1}{2}}\,\varepsilon \tag{20}$$

and if the ellipses are strongly eccentric (e. g. $\beta_1 = 0.1, \beta_2 = 10$) we end up with $\varepsilon_p \cong 10\varepsilon$.

However, albeit the slices have the same number of electrons and the same emittance, they do not contribute to the laser intensity in the same way. Since the $\rho$ parameter depends on the current density, we find that the slice gain parameter behaves like $\rho \propto \beta^{-1/3}$ ‡, we find therefore that the slice with smaller $\beta$ has smaller saturation length. The slice with larger $\beta$ may even not reach the saturation. It could also happen that a slice, characterized by a large *c*-coefficient and by a large value of $\beta$, may have the same saturation length of a slice with a smaller number of electrons but with smaller $\beta$, provided that $c_1\beta_2 = c_2\beta_1$. In this case even though the individual slice emittances are small the effective emittance of the bunch becomes that shown in Fig. 4. This means that the brightness of the emitted radiation could be significantly diluted by this effective emittance.

A better idea of the interplay between e-beam transport and slice phase space dynamics is given by Fig. 6, where we have the evolution along the SPARC channel of the slices Twiss coefficients, of the e-beam transverse section and of the Courant Snyder ellipses on the successive diagnostic flags.

## III. PHASE SPACE EVOLUTION AND FEL DYNAMICS

In the previous sections we have assumed that the phase space of the projected emittance is essentially Gaussian, this is just an approximation which allows to define an average projected emittance and average Twiss coefficients.

There is no reason that the distribution given in eq. (5) be Gaussian, this is evident from Figs. 7, where we have reported the spatial and angular distribution for the projected distribution and for a single slice, whose individual phase space contour plots are reported in Fig. 8.

The deviation from a Gaussian can be determined by noting that the even higher order moments of a Gaussian satisfy the relation

$$\left\langle x^{2m}\right\rangle_n = \int_{-\infty}^{\infty} dx' \int_{-\infty}^{\infty} dx\, x^{2m} f_n(x,x') = \frac{(2m)!}{m!}\left(\frac{\sigma_{x,n}}{\sqrt{2}}\right)^{2m}, \qquad (21)$$
$$\sigma_{x,n} = \sqrt{\beta_n \varepsilon_n}$$

we define, therefore, the following parameter §

$$Q_m = \frac{\left\langle x^{2m}\right\rangle_p - \frac{(2m)!}{m!}\left(\frac{\sqrt{\left\langle x^2\right\rangle_p}}{\sqrt{2}}\right)^{2m}}{\frac{(2m)!}{m!}\left(\frac{\sqrt{\left\langle x^2\right\rangle_p}}{\sqrt{2}}\right)^{2m}} \tag{22}$$

where the subscript $p$ denotes that the average has been taken over the projected distribution as a measure of such a deviation.

The parameter $Q_m$ for the cases relevant to Figs. 7,8 is shown in Fig. 9, which indicates a significant deviation from the Gaussian distribution.

The preliminary conclusion we may draw from the discussion developed in this note is that the slice emittance alone is not sufficient to specify the FEL SASE spike competition and that a significant improvement may come from the understanding of the evolution of the relevant Twiss parameters during the evolution. This aspect of the problem can be understood by noting that the small signal high gain FEL equation should be written as [4]

$$\partial_\tau a(\tau) = i\pi g_0 \sum_n c_n \int_0^\tau d\tau' \frac{e^{-i\nu\tau'}}{R_{x,n}(\tau')R_{y,n}(\tau')} a(\tau-\tau') \tag{23}$$

where

$$R_{\eta,n}(\tau) = \sqrt{(1+\alpha_{\eta,n}^2)\left(1-i\pi\mu_{\eta,n}\tau\right)(1-i\pi\mu'_{\eta,n}\tau)-\alpha_{\eta,n}^2},$$

$$\mu_{\eta,n} = \frac{4N\gamma^2\varepsilon_{\eta,n}}{(1+\frac{K^2}{2})\gamma_{\eta,n}}k_{\beta,n}^2, \mu'_{\eta,n} = \frac{4N\gamma^2\varepsilon_{\eta,n}}{(1+\frac{K^2}{2})\beta_{\eta,n}}, \tag{24}$$

$k_\beta = betatron\ motion\ wave\ number$

We have so far just fixed the terms of the problem and we will now discuss a specific case. We have sampled the simulated distribution of the SPARC e-beam (Table I) at the entrance of the undulator in 12 slices with a length of 288 μm each, as shown in Figs. 10a and 10b reporting respectively the slices current and the relevant slices Twiss parameter and emittances.

The procedure we follow to provide a very quick evaluation of the individual slice evolution is that of expressing the evolution of the SASE power in terms of the following logistic equation [6]

$$P(z_b,z)=\sum_n h_n(z_b)\cdot P_0\cdot\frac{A_n(z)}{1+\frac{P_0}{P_{F,n}(z)}\cdot\left(A_n(z)-1\right)}$$

$$A_n(z)=\frac{1}{9}\left(3+2\cosh\left(\frac{z}{L_{g,n}(z)}\right)+4\cos\left(\frac{\sqrt{3}}{2}\frac{z}{L_{g,n}(z)}\right)\cdot\cosh\left(\frac{z}{2L_{g,n}(z)}\right)\right) \quad (25).$$

which is an incoherent sum of the different logistic functions yielding the evolution of each slice. The index *n* refers to the slice number and the gain lengths for each slice have been evaluated using the emittance values reported in Fig. 10b, furthermore the evolution of the Twiss coefficients has been included too, by taking into account the relevant evolution along the transport channel, as shown in Fig. 11.

The results of the slice evolution, including slippage effects too, along the electron bunch coordinate at different position inside the undulator, are shown in Figs. 12, 13.

The plots reflect what we expect on physical grounds, namely the dominant slice is selected by the combination of different effects due to the emittances, current, matching an so on.

A more global idea is offered by Fig. 14**,** where we have reported the integrated power evolution of the first, seventh and tenth slice. It is evident that the first slice has no chance of reaching the saturation.

We have not considered the spiking dynamics, which can however be accounted for too, using this simplified procedure.

In a forthcoming investigation we will treat the transverse mode dynamics by taking into account the full transverse phase space dynamics.

## IV CONCLUDING REMARKS

It is now worth going back to our main assumption that the slice phase space transverse distribution is given by eq. (4), according to which the phase space ellipse centroids are all located at the origin. This is not a-priori ensured [7], therefore it could be more appropriate to consider the following distribution

$$f_n(x,x') = \frac{1}{2\pi\varepsilon_n}\exp\left[-\frac{\gamma_n(x-\xi_n)^2 + 2\alpha_n(x-\xi_n)(x'-\xi_n^{'}) + \beta_n(x'-\xi_n^{'})^2}{2\varepsilon_n}\right] \qquad (26)$$

where $\xi_n, \xi_n^{'}$ are the phase space centers of the individual slices Courant Snyder ellipses.

This effect can be the source of a further increase of the projected emittance, the situation is better illustrated by Fig. 15a, in which we have reported the different slice phase space distribution and its projected counterpart.

The effect of the centre mismatch on the spatial distribution is even more interesting due to the possible appearance of peaks (see Fig. 15b).

To better quantify this effect we can proceed in the following way: the average and rms values of the projected phase space distribution will be defined by

$$\Xi = \sum_{n=1}^{m} c_n \xi_n, \Xi' = \sum_{n=1}^{m} c_n \xi_n^{'},$$

$$\sigma_x^2 = \left\langle x^2 \right\rangle - \left\langle x \right\rangle^2 = \sum_{n=1}^{m} c_n \beta_n \varepsilon_n + \Delta,$$

$$\sigma_{x'}^2 = \left\langle x'^2 \right\rangle - \left\langle x' \right\rangle^2 = \sum_{n=1}^{m} c_n \gamma_n \varepsilon_n + \Gamma,$$

$$\sigma_{x,x'} = \left\langle x\, x' \right\rangle - \left\langle x \right\rangle \left\langle x' \right\rangle = \sum_{n=1}^{m} c_n \alpha_n \varepsilon_n + \Lambda, \qquad (27).$$

$$\Delta = \sum_{n=1}^{m} c_n \xi_n^2 - \Xi^2, \Gamma = \sum_{n=0}^{m} c_n \xi_n^{'2} - \Xi^{'2},$$

$$\Lambda = \sum_{n=1}^{m} c_n \xi_n^{'} \xi_n - \Xi \Xi'$$

According to the previous relation we obtain, for the projected emittance the following expression

$$\varepsilon_p^* = \varepsilon_p + \varepsilon_c + \varepsilon_M,$$

$$\varepsilon_c = \Delta\Gamma - \Lambda^2, \qquad (28).$$

$$\varepsilon_M = \sum_{n=1}^{m} c_n \varepsilon_n \left[ \beta_n \Gamma + \gamma_n \Delta - 2\alpha_n \Lambda \right]$$

In which one can identify two extra contributions with respect to the value calculated with eq. (9).

A different strategy, less mathematically cumbersome, could be that of defining a phase space area of the slice centroids.

We could indeed consider the following quantities

$$\sigma_{\xi}^{2}=\left\langle\xi^{2}\right\rangle-\left\langle\xi\right\rangle^{2},$$
$$\sigma_{\xi'}^{2}=\left\langle\xi'^{2}\right\rangle-\left\langle\xi'\right\rangle^{2},$$
$$\sigma_{\xi,\xi'}=\left\langle\xi\xi'\right\rangle-\left\langle\xi\right\rangle\left\langle\xi'\right\rangle, \qquad (29)$$
$$\left\langle a\right\rangle=\frac{1}{m}\sum_{n=1}^{m}a_{n}$$

thus introducing the centroid emittance and Twiss coefficients, namely

$$\varepsilon_{\xi}=\sqrt{\sigma_{\xi}^{2}\sigma_{\xi'}^{2}-\sigma_{\xi,\xi'}^{2}},$$
$$\beta_{\xi}=\frac{\sigma_{\xi}^{2}}{\varepsilon_{\xi}},\gamma_{\xi}=\frac{\sigma_{\xi'}^{2}}{\varepsilon_{\xi}},\alpha_{\xi}=\frac{\sigma_{\xi,\xi'}}{\varepsilon_{\xi}} \qquad (30).$$

We can define the associated phase space distribution and specify the projected distribution as a convolution between the two having the momenta given by eqs. (29) and (9). In this way the centroids contribute to the projected emittance in a kind of diffusive way. The centroid spreading can be due to various mechanisms, which will be carefully discussed in a forthcoming investigation.

Here we want to stress the relevance of the slice energy spread. If any slice is characterized by a longitudinal energy distribution of the type

$$\varphi_{n}(z,\varepsilon)=\frac{1}{2\pi\Sigma_{\varepsilon,n}}\exp\left(-\frac{\beta_{\varepsilon,n}\varepsilon^{2}+\gamma_{\varepsilon,n}z^{2}}{2\Sigma_{\varepsilon,n}}\right) \qquad (31)$$

with $\Sigma_{\varepsilon,n}$ being the longitudinal emittance and with

$$\varepsilon = \frac{\gamma - \gamma_0}{\gamma_0} \tag{32}$$

being the relative energy. The single slice energy spread and bunch length can therefore be defined as

$$\begin{aligned} \sigma_{\varepsilon,n} &= \sqrt{\gamma_{\varepsilon,n}\Sigma_{\varepsilon,n}}, \\ \sigma_{z,n} &= \sqrt{\beta_{\varepsilon,n}\Sigma_{\varepsilon,n}} \end{aligned} \tag{33}$$

The “chromatic” structure of the packet will reflect itself into the magnification of the chromatic effects inside transport elements like quadrupoles and solenoids, any individual slice will indeed be affected in a different way by the energy dependent part of the transport element.

An example may be provided by a solenoid misalignments [8], causing a vertical component of the magnetic field, which will induce a coupling of the motion in the $x - z$ plane, which will be characterized by centroid shift, which can be expressed as

$$\begin{aligned} \xi_n &\propto \Lambda^{-1}\sigma_{\varepsilon,n}, \\ \Lambda &= \frac{m_0\gamma_0 c}{e\,\delta B} \end{aligned} \tag{34}$$

The distribution given in eq. (31) does not contain any energy phase correlation term, whose effect will be discussed elsewhere along with its important role in the physics of FEL.

## V. ACKNOWLEDGMENTS

The Authors express their sincere appreciation to Drs. F. Ciocci and L. Giannessi for interesting discussions and suggestions and for providing the plots of Fig. 6. It is also a pleasure to thank Drs. P.L. Ottaviani and S. Pagnutti for running the code Prometeo. One of the Authors (G.D.) acknowledges enlightening discussions with Drs. M. Ferrario and D. Filippetto on the effect of the phase space centroid mismatches

## FOOTNOTES AND REFERENCES

* Such a definition holds if $l_c << \sigma_z$ otherwise it should be replaced by an integral.

† $\hat{A}$ and $\underline{x}$ are an $n \times n$ matrix and an $n$ components vector respectively.

‡ We have assumed a round slice with the same vertical and horizontal beta and emittances.

§ This quantity is a generalization of the Mandel $Q$ parameter, used in quantum optics to study the statistical properties of radiation.

**FIGURE CAPTIONS**

Fig 1. Optical field distribution at saturation for the SPARC operation (continuous line) and electron bunch distribution (dot line)

Fig 2. a) Slice sampling of a Gaussian bunch; b) transverse sections of the sliced bunch

Fig 3. Courant Snyder Ellipse and Twiss parameters

Fig 4. Courant Snyder Ellipses a) Slice phase space b) Projected phase space

Fig 5. Courant Snyder Ellipses for two individual slices (blue and red) and for the projected emittance (cyan)

Fig 6. Slice transport along the SPARC magnetic channel, the Twiss parameters and the emittances of each slice have been generated randomly and the matching has been provided for the central slice denoted by a blue curve in the upper figure, where we have reported the (radial) beta function evolution of each slice. The lower curves, referring to the beam diagnostic flags along the channel, provide the beam section and the radial phase space evolution

Fig 7. a) Spatial transverse distribution (slice green, projected red); b) angular distribution (slice green, projected red)

Fig 8. Courant Snyder ellipses for the different slices (the green curves in Fig. 7 refers to the green ellipse)

Fig 9. $Q_m$ parameter for the first 7 moments of the projected spatial distribution, for spatial (squares) and angular (round)

Fig 10. a) SPARC electron bunch and associated sample slice and currents; b) slice Twiss parameters and emittances at the entrance of the undulator

Fig 11. Evolution along the transport Channel of the $\beta$ Twiss parameter of the individual and projected slices

Fig 12. Slice evolution at the beginning of the undulator

Fig 13. Slice evolution at the middle of the undulator

Fig 14. Integrated power evolution vs. the undulator length of the first (solid), seventh (dash) and tenth (dot) slice

Fig 15. a) The contribution to the increase of the projected emittance of the phase space centre mismatches of the individual slices (the slash contour in the plot on the right refers to projected emittance for slices having all the same phase space centre); b) spatial distribution of two slices (brown, blue) and projected (red)

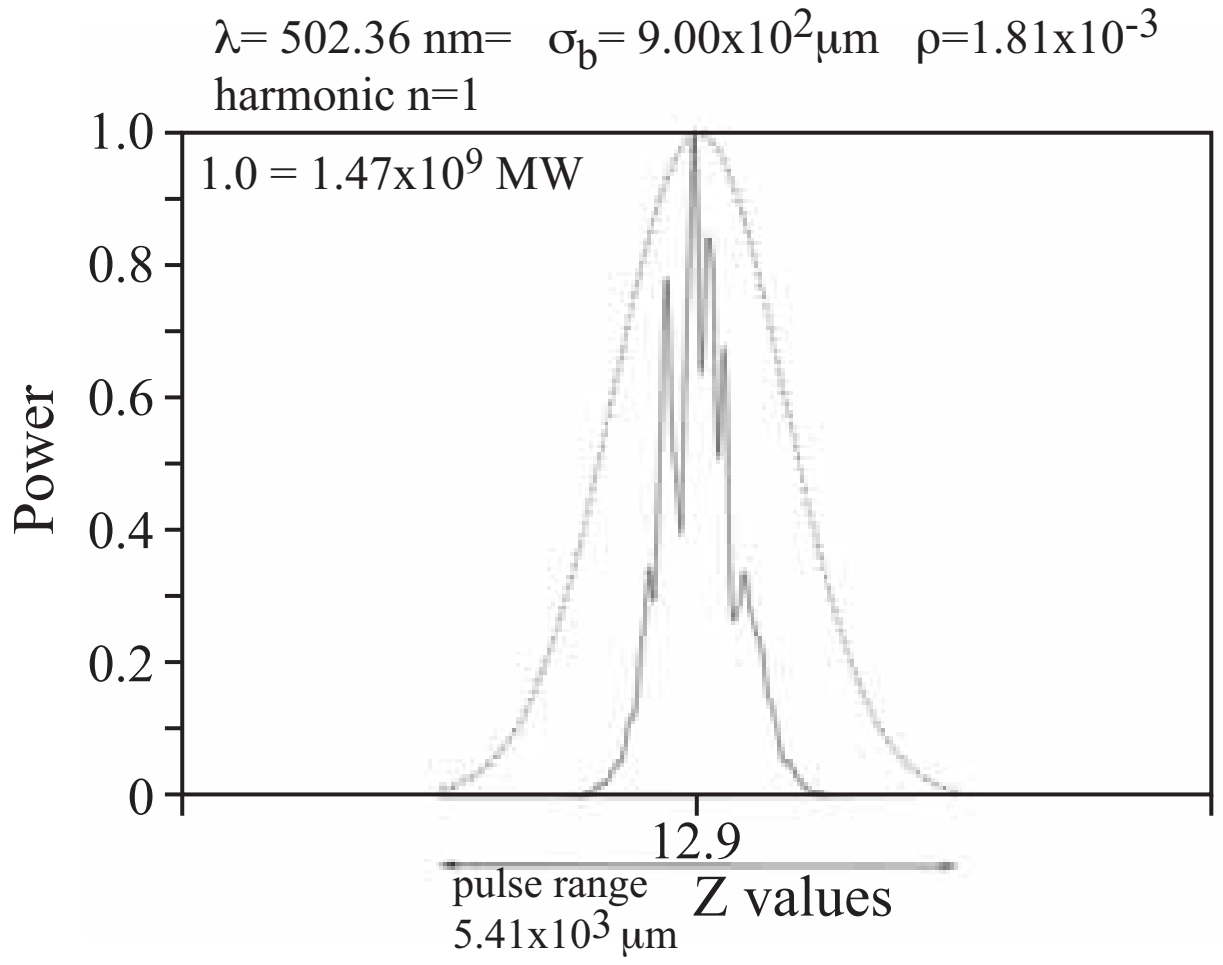

Fig.1

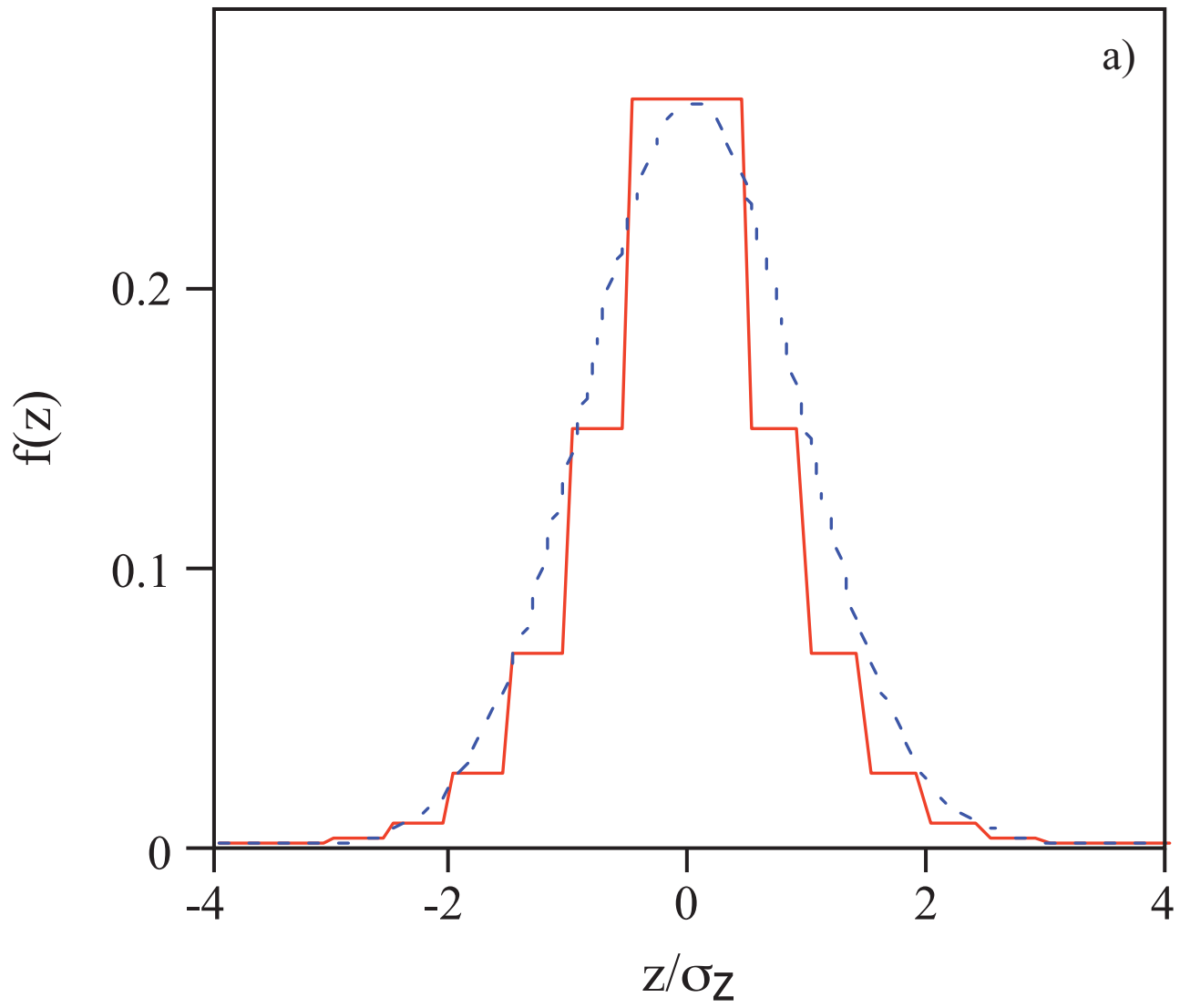

Fig2a

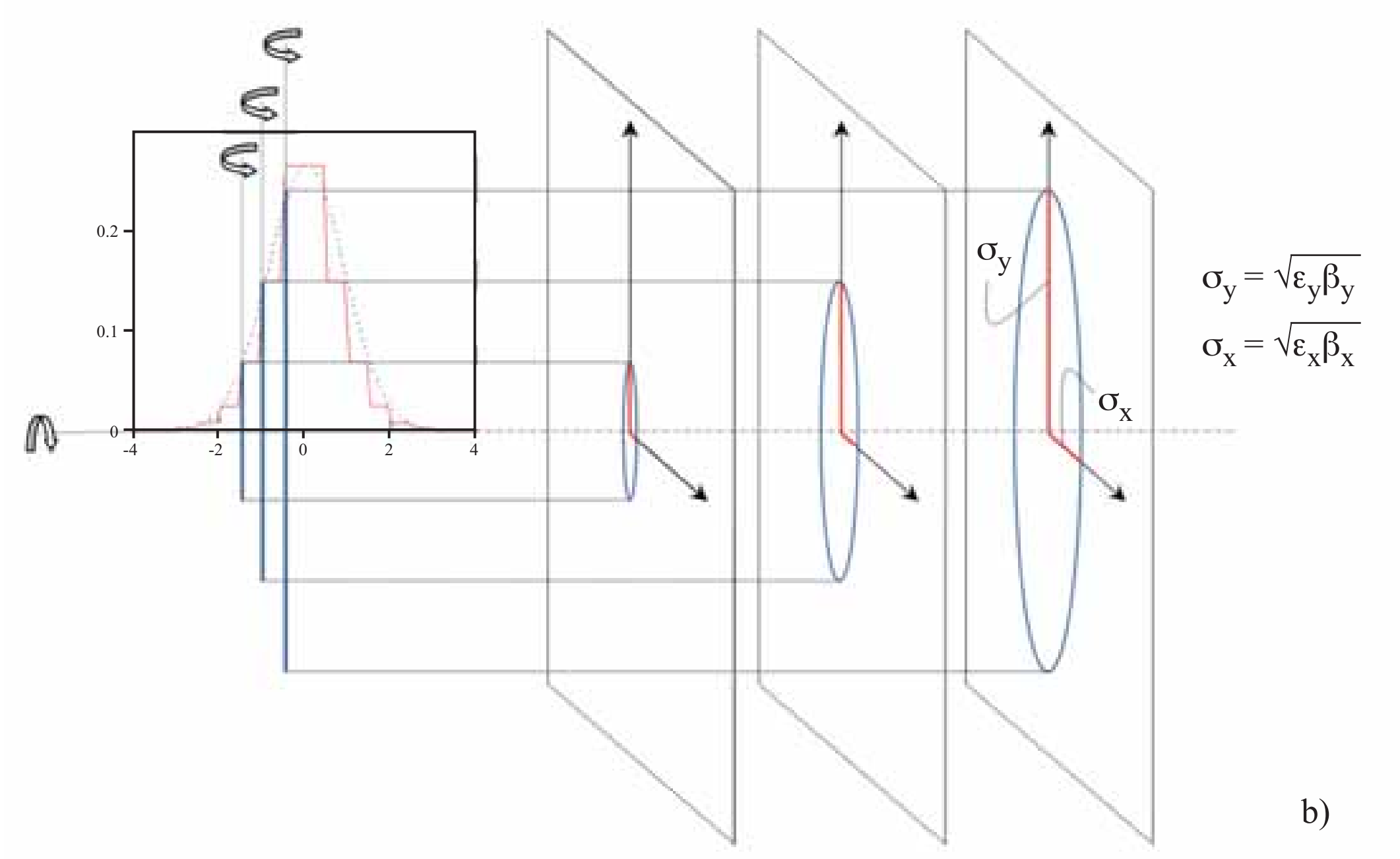

Fig.2b

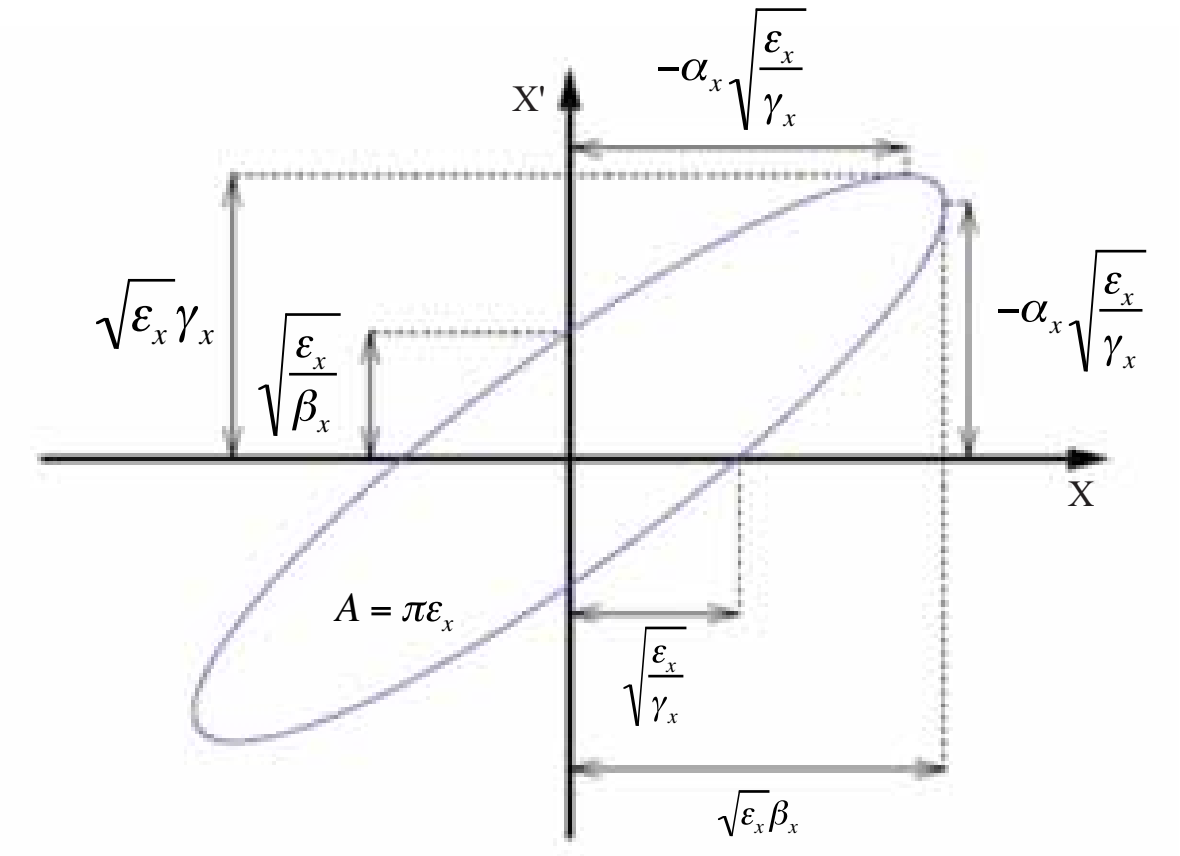

Fig3

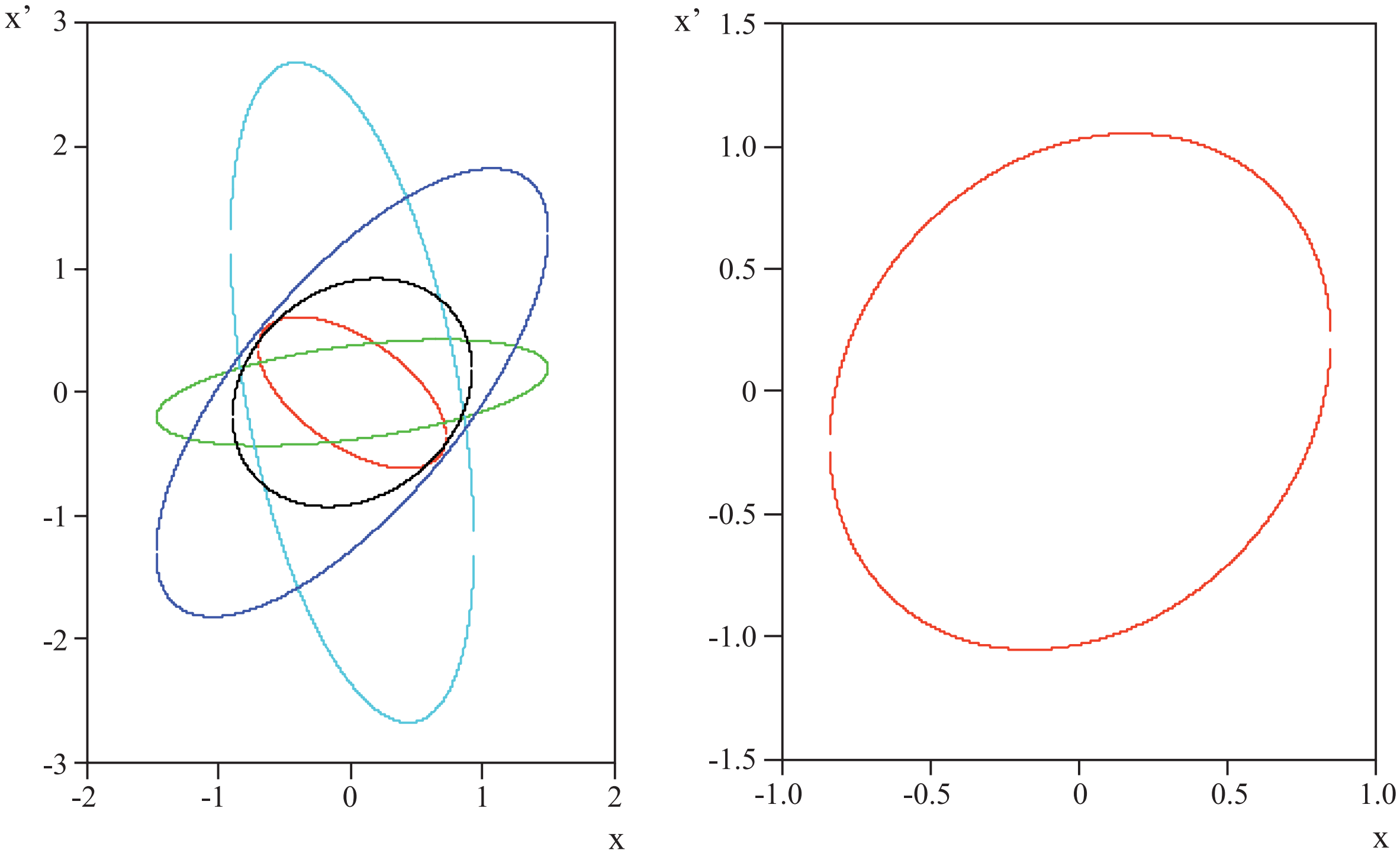

Fig.4

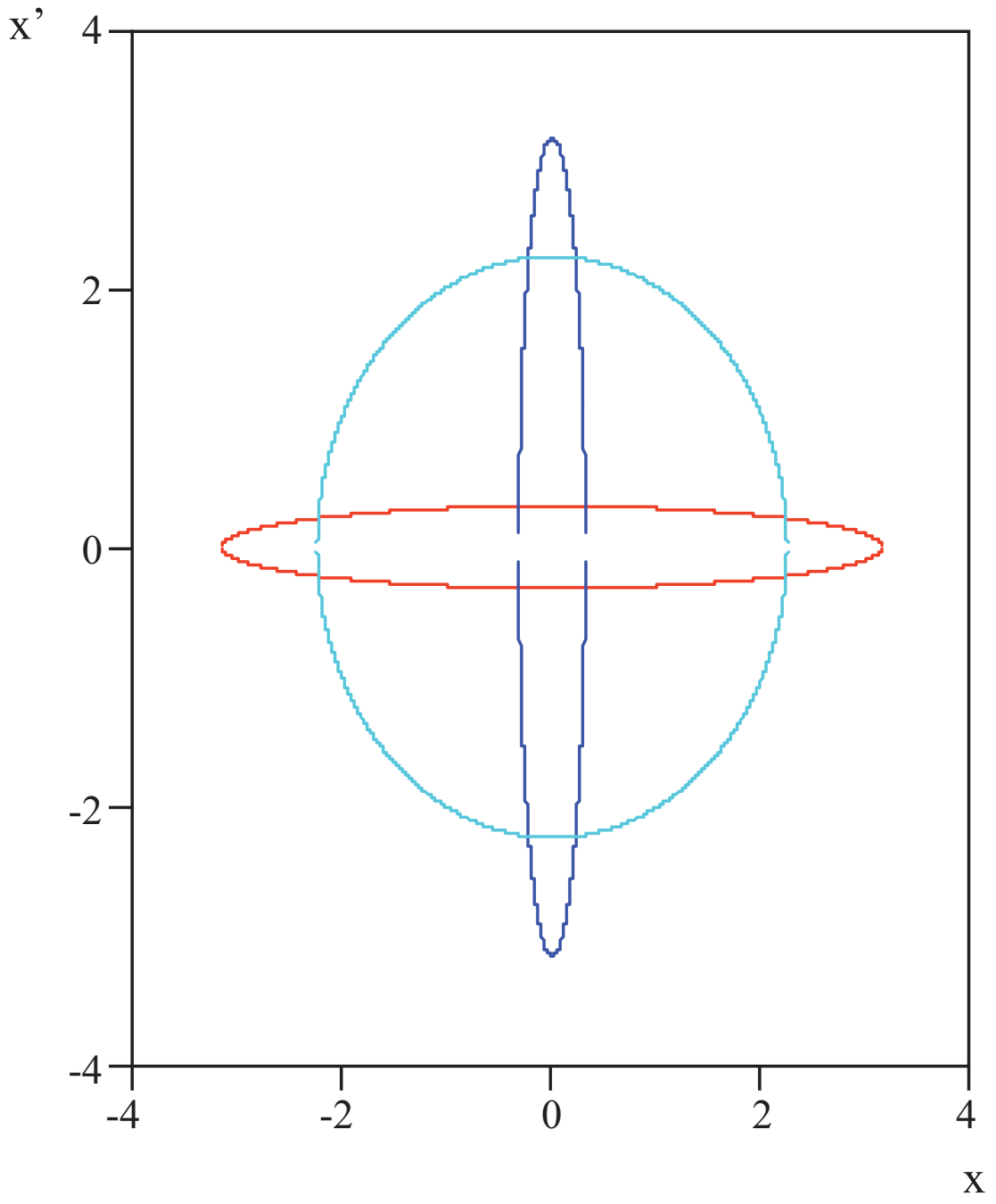

Fig5

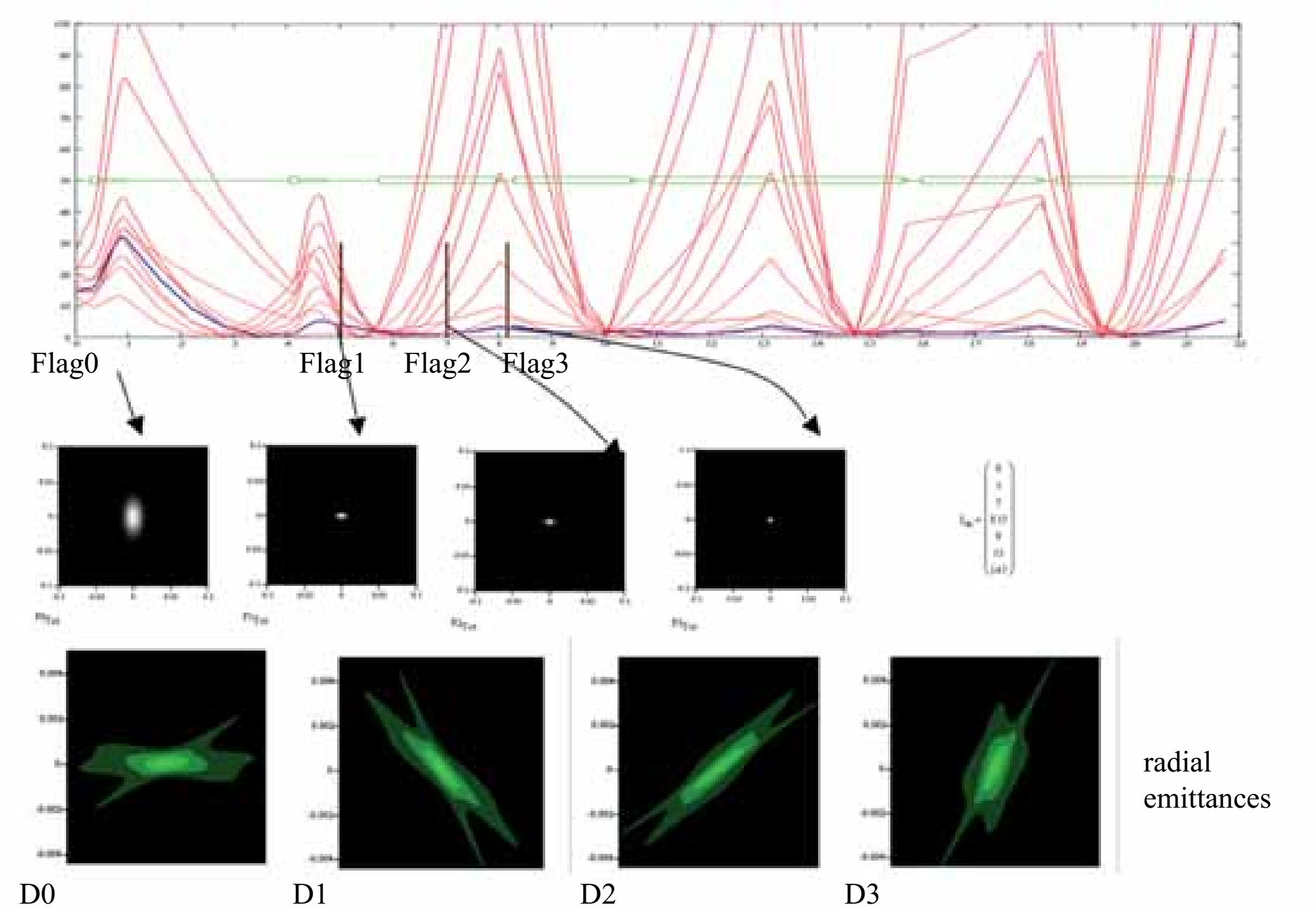

Fig.6

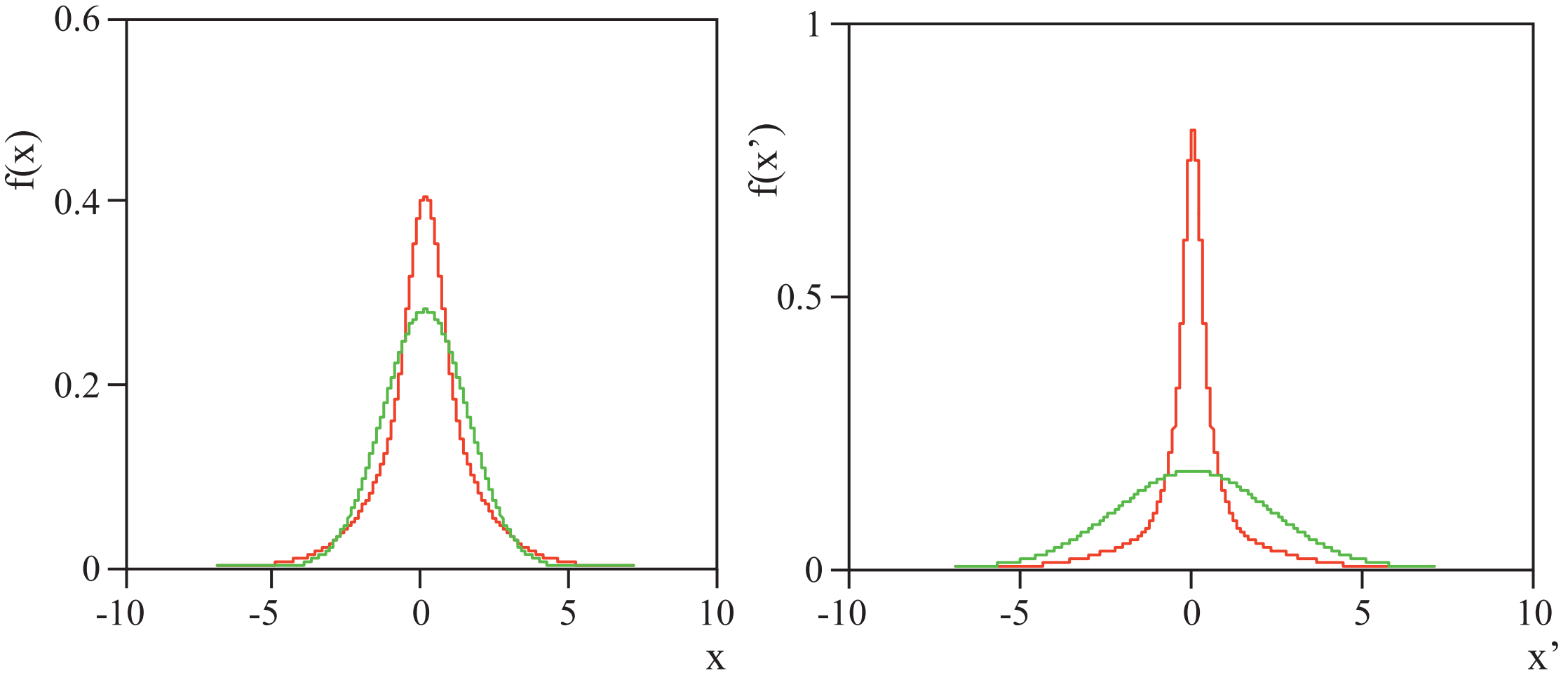

Fig7

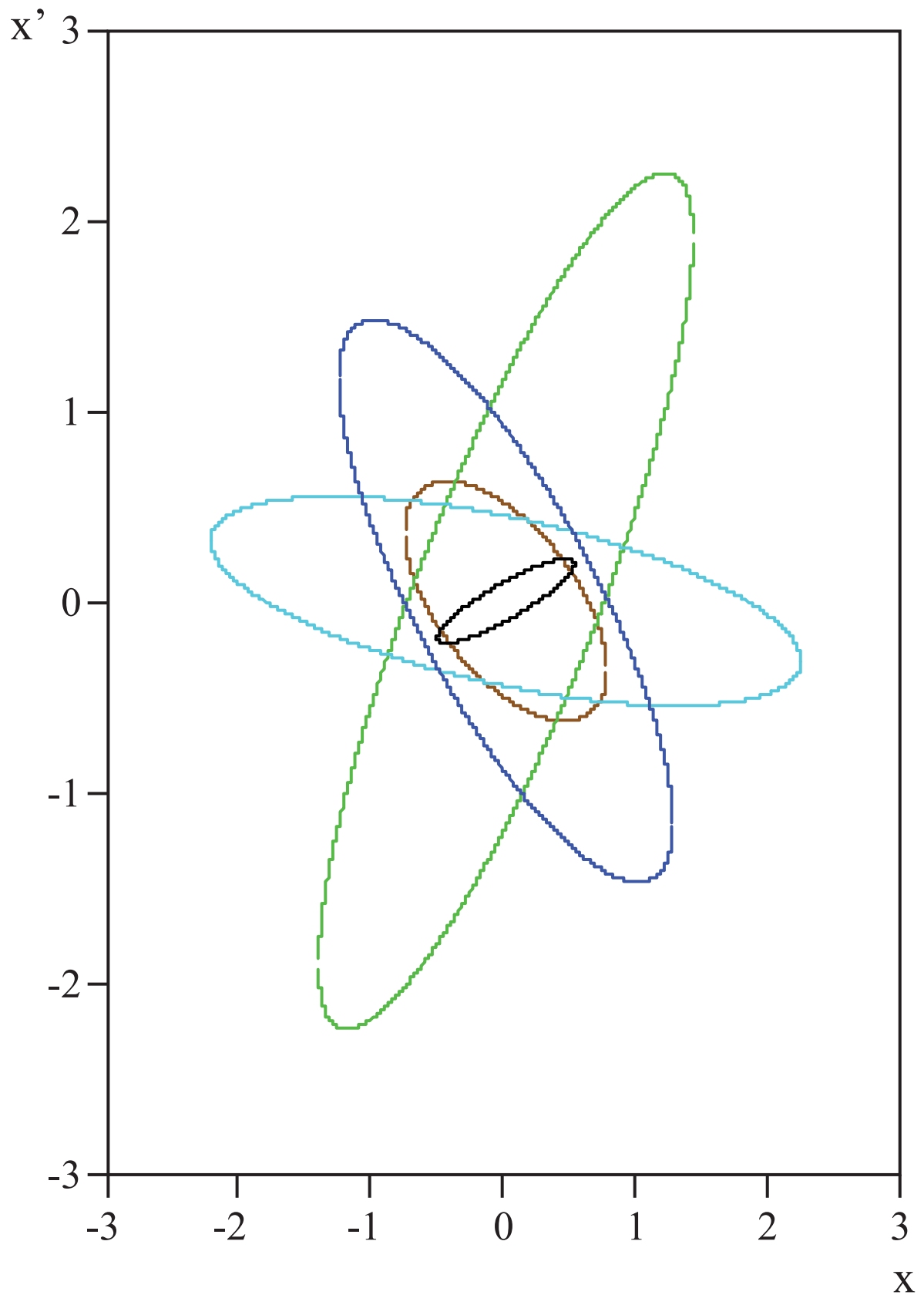

Fig8

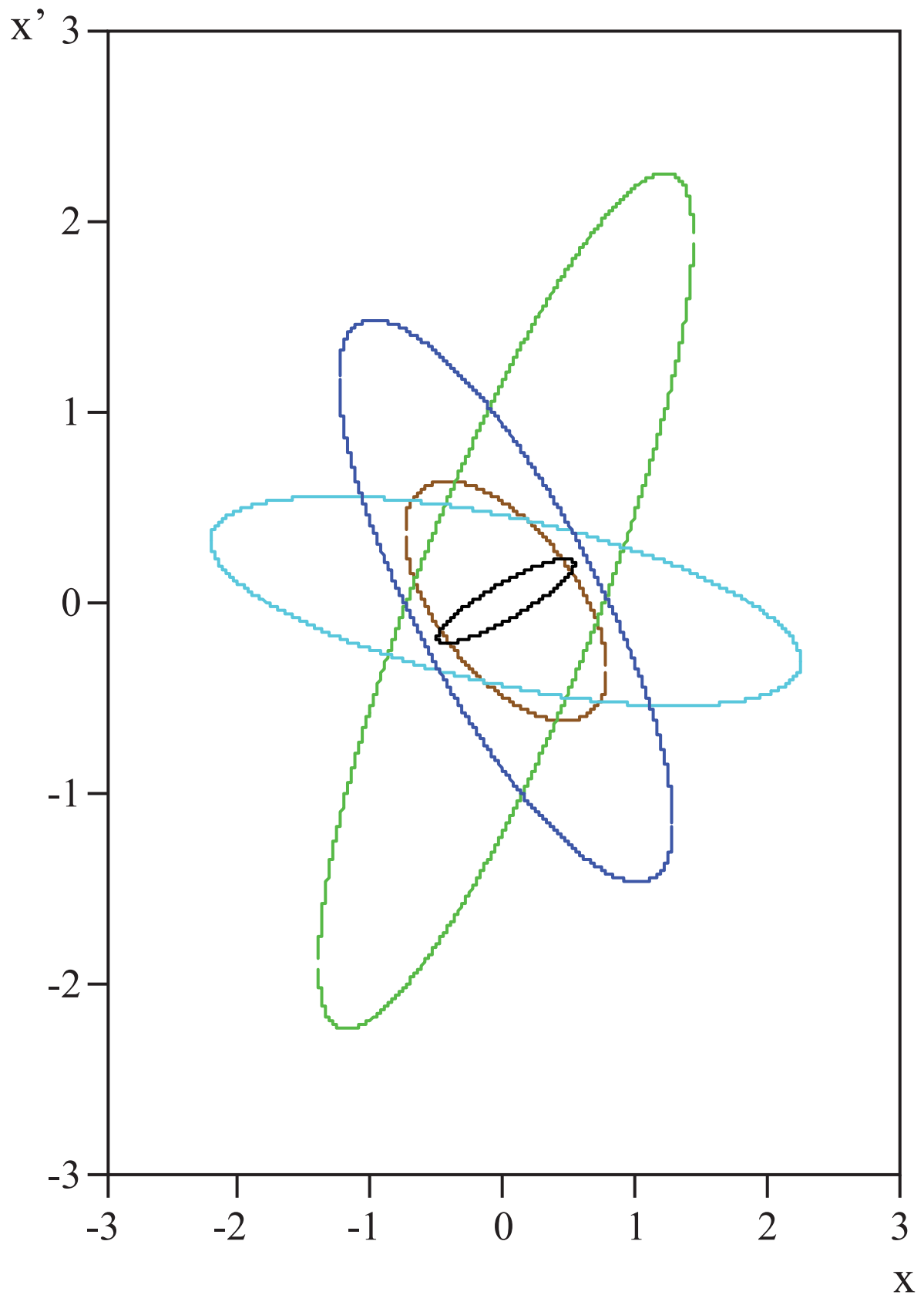

Fig.9

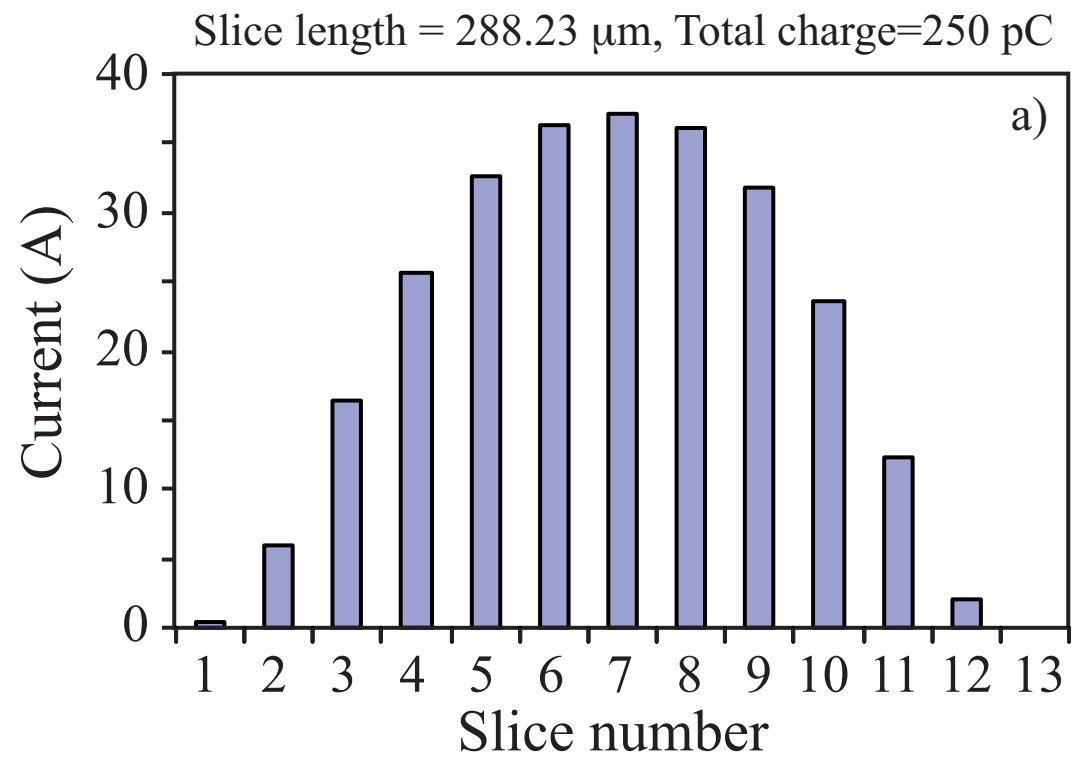

Slice length = 288.23 µm, Total charge=250 pC
a)
Current (A)
40
30
20
10
0
1 2 3 4 5 6 7 8 9 10 11 12 13
Slice number

Fig.10a

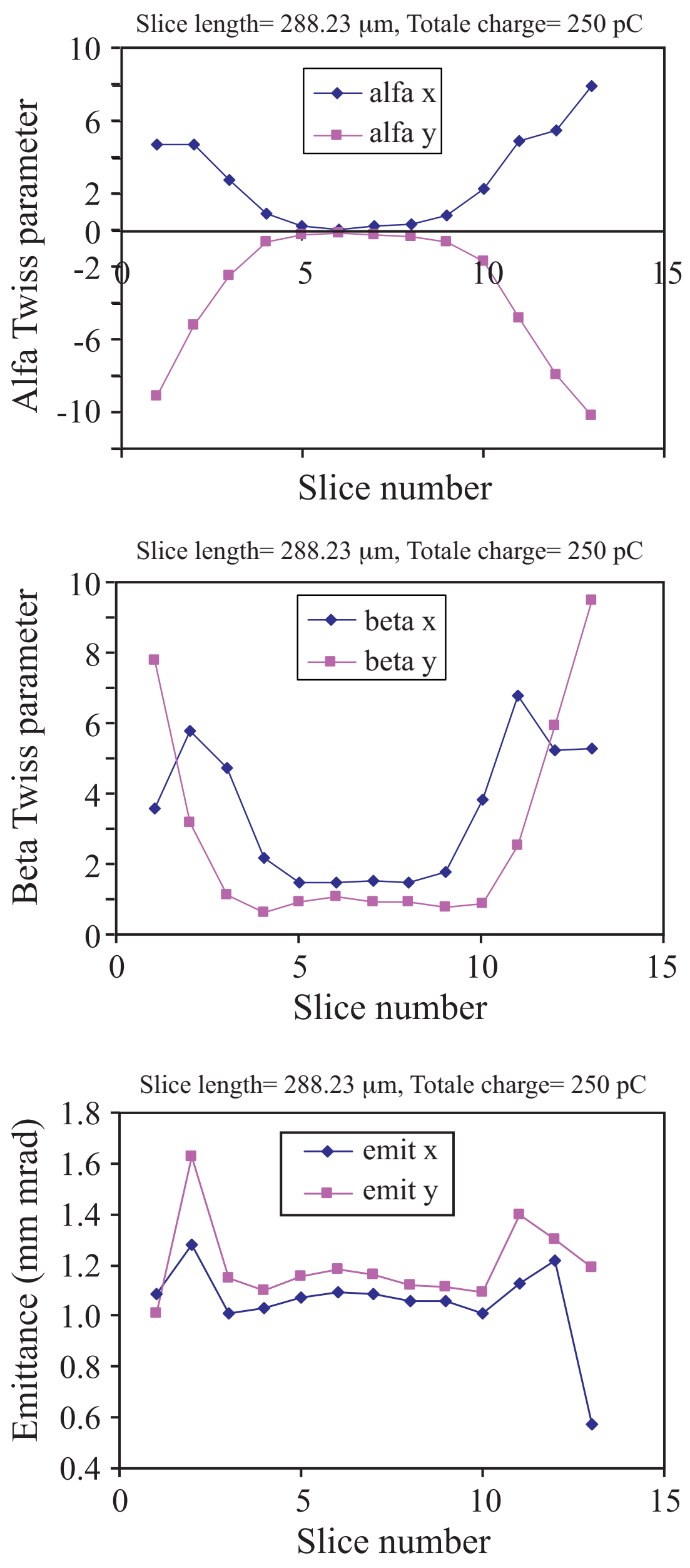
Slice length= 288.23 μm, Totale charge= 250 pC
alfa x
alfa y
Alfa Twiss parameter
Slice number
Slice length= 288.23 μm, Totale charge= 250 pC
beta x
beta y
Beta Twiss parameter
Slice number
Slice length= 288.23 μm, Totale charge= 250 pC
emit x
emit y
Emittance (mm mrad)
Slice number

Fig.10 (123)

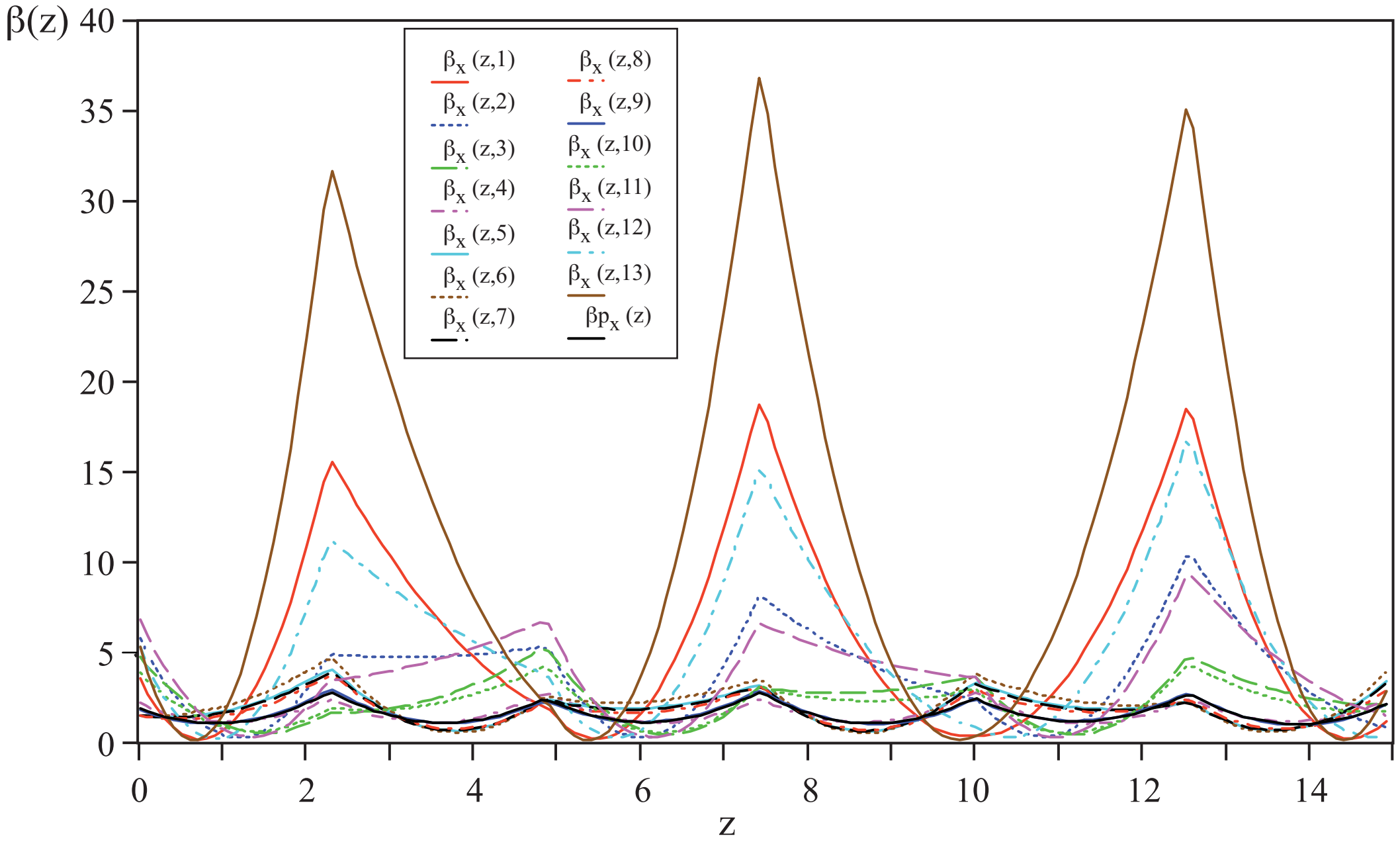

Fig.11

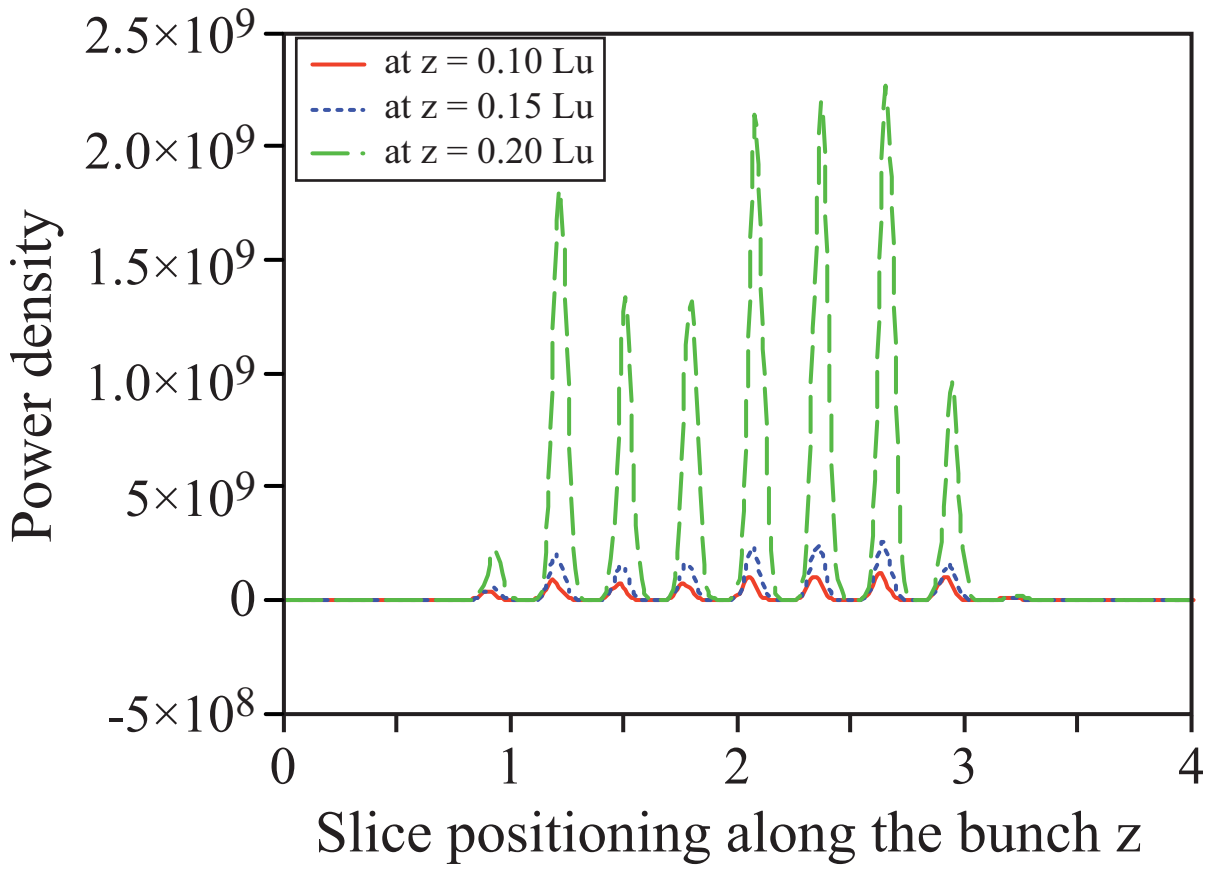
at z = 0.10 Lu
at z = 0.15 Lu
at z = 0.20 Lu
Power density
Slice positioning along the bunch z

Fig.12

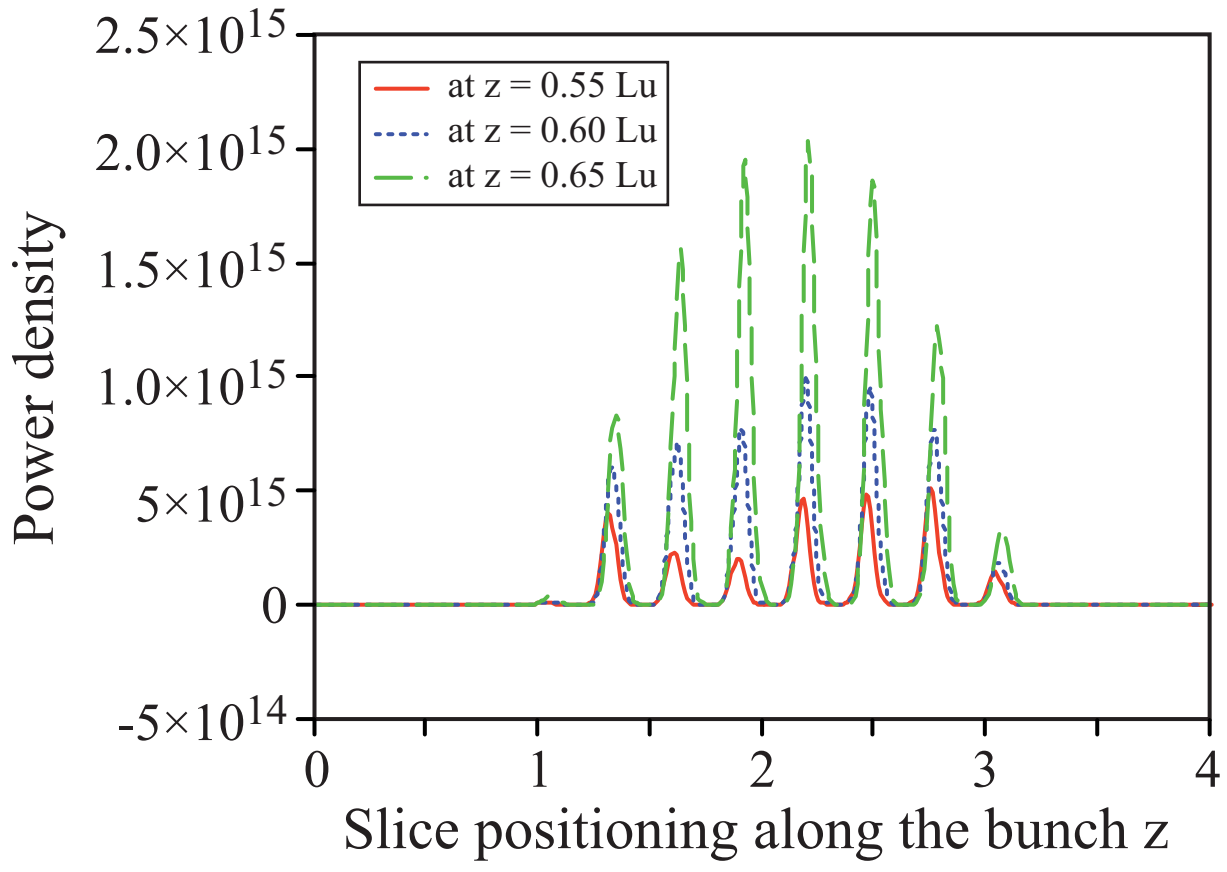

2.5×10^15
2.0×10^15
1.5×10^15
1.0×10^15
5×10^15
0
-5×10^14
at z = 0.55 Lu
at z = 0.60 Lu
at z = 0.65 Lu
Power density
0
1
2
3
4
Slice positioning along the bunch z

Fig.13

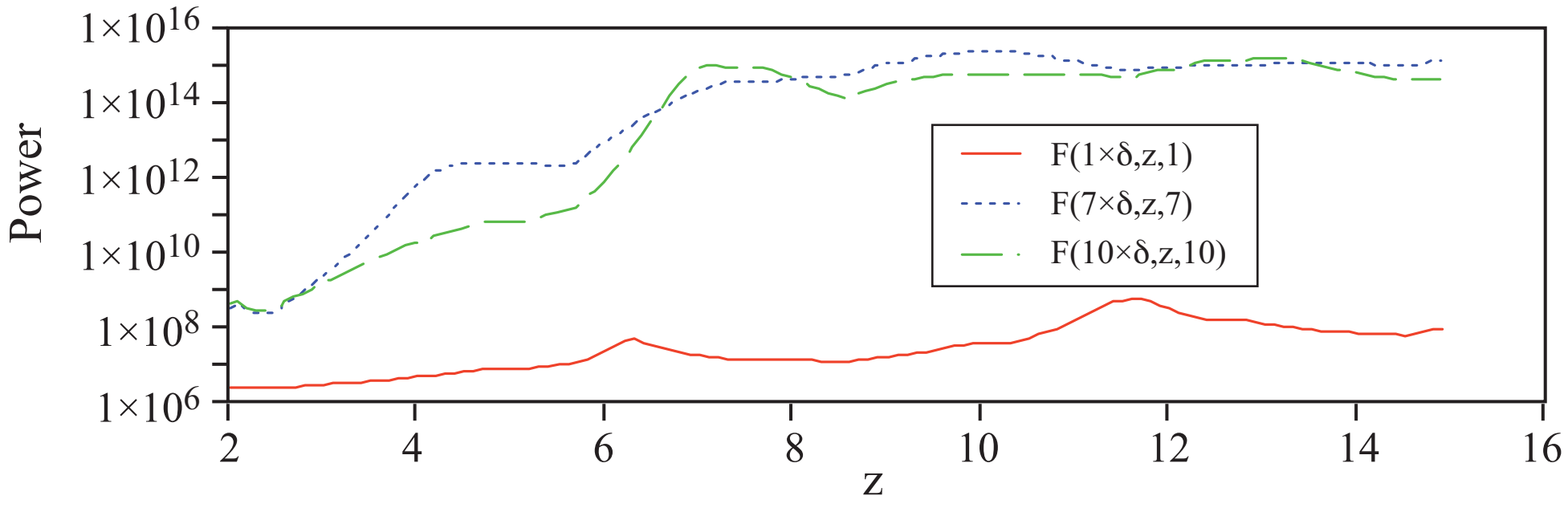

Fig.14

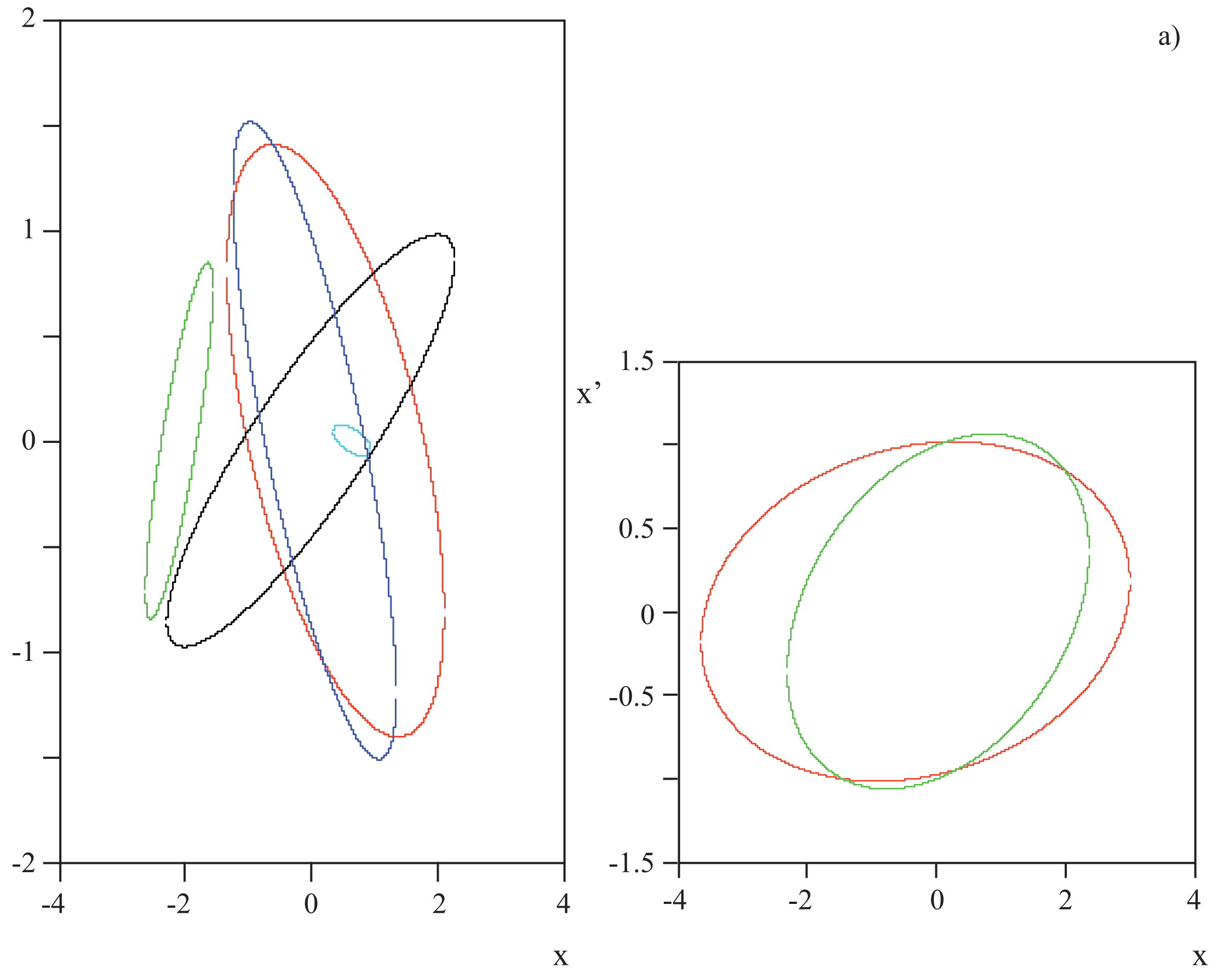

Fig.15a

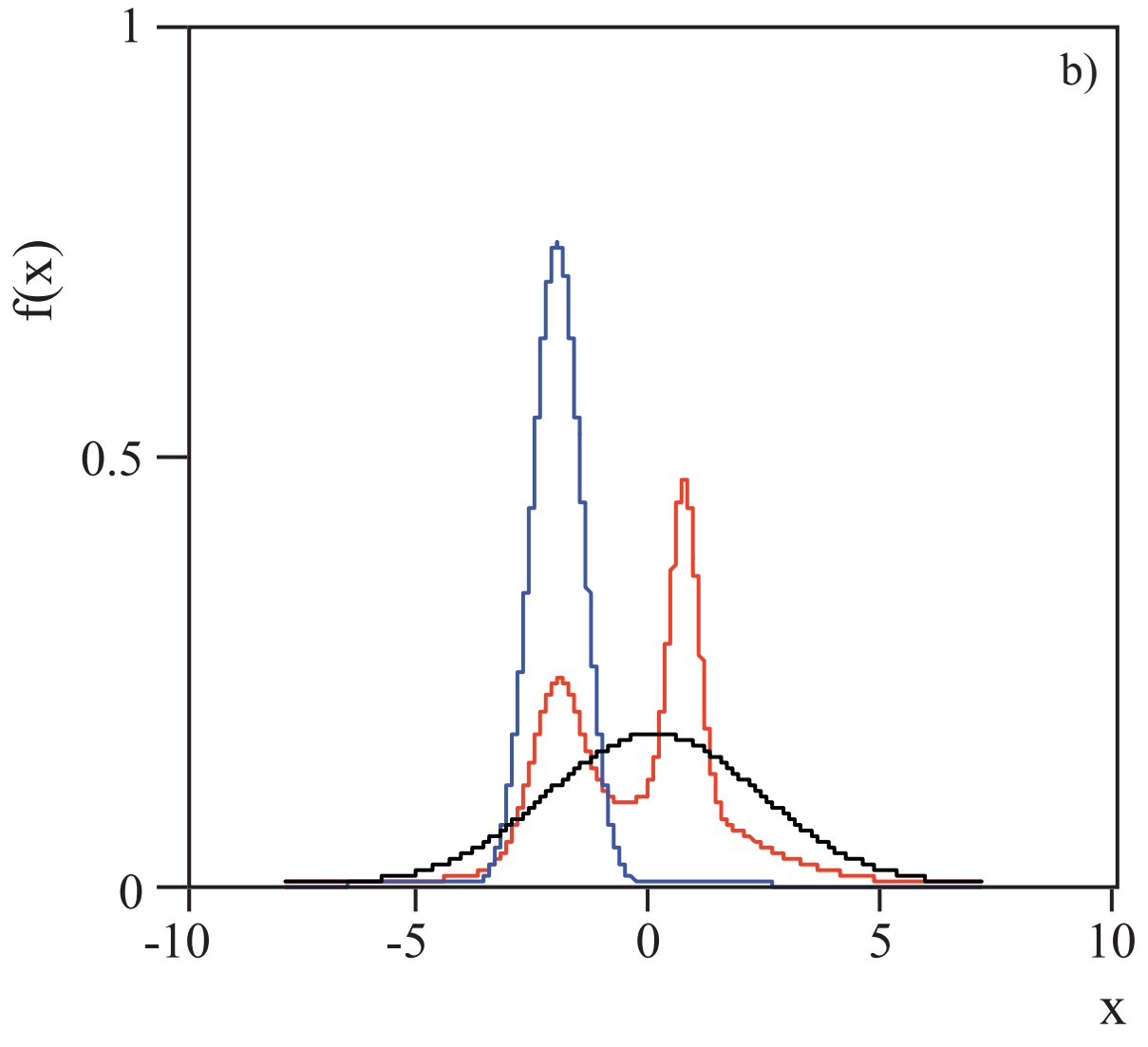

Fig.15b

**Table I - Global SPARC beam parameters**

| | |
|---|---|
| Energy | 146 MeV |
| Charge | 250 pC |
| Projected horizontal emittance | 1.4 mm-mrad |
| Projected vertical emittance | 1.61 mm-mrad |
| Global Twiss parameters | 〈x=0.886, ®x=1.868, 〈y=-0.7581, ®y=0.789 |